# Formation of Oriented Bilayer Motif: Vanadyl Phthalocyanine on Ag(100)

W. Koll[1], C. Urdaniz[2,3], Kyungju Noh[2,4], Y. Bae[2,4], C. Wolf[2,4], J.A. Gupta[1]

[1] Department of Physics, Ohio State University
[2] Center for Quantum Nanoscience, Institute for Basic Science (IBS), South Korea
[3] Ewha Womans University, South Korea
[4] Department of Physics, Ewha Womans University, South Korea

The adsorption and self-assembly of vanadyl phthalocyanine molecules on Ag(100) has been investigated using a combination of scanning tunneling microscopy and density functional theory. At sub-monolayer coverage, we observe two distinct adsorption configurations of isolated molecules, corresponding to the central O atom pointing toward ("O-down") or away ("O-up") from the substrate. Upon adsorption in the O-up orientation, the otherwise achiral molecules take on a windmill-like chiral appearance due to their interaction with the substrate. At monolayer coverage, we observe a self-assembled square lattice with a mixture of O-up and O-down molecules. At higher coverage we find a strong preference for bilayer formation with O-up and O-down molecules in alternating layers, suggesting stabilization by dipolar interactions. Close inspection of the multi-layer surface reveals grain boundaries separating domains of opposite organizational chirality, and long-range ordering.

**Introduction**

Metal phthalocyanine (MPc) molecules have been the subject of extensive research owing to their desirable electrical, magnetic, optical, and catalytic properties.[1–6] As a result, they have found applications in photovoltaics, light-emitting diodes, and field-effect transistors.[1,3,4] Incorporation of MPc molecules in such electrical devices requires an understanding of their functionalization with inorganic interfaces. This motivates scanning tunneling microscopy (STM) studies of MPc molecules to probe their interactions with a variety of surfaces.[2] Despite being intrinsically achiral in the gas phase, STM imaging shows that chirality is induced in several MPcs upon adsorption on surfaces.[7–13] Self-assembly of these molecules leads to an 'organizational' chirality as well, favored by intermolecular interactions and substrate registry.[14] More recent STM studies have focused on nonplanar MPcs such as TiOPc,[15] ClAlPc,[16] PbPc,[17] and SnPc,[18] whose intrinsic dipole moments modify intermolecular and surface interactions compared to their planar counterparts. For example, multilayer films of nonplanar MPcs adopt a bilayer motif, wherein dipole moments of molecules in subsequent layers alternate directions.[15–18] Adsorption-induced chirality has also been reported for a variety of nonplanar MPc.[19–23] In the nonplanar MPc family, the spin $S=½$ vanadyl phthalocyanine (VOPc) has gained attention as a molecular qubit candidate with a long spin lifetime persisting up to room temperature.[24] Most STM studies of VOPc have focused on lower coverage regimes where substrate interactions are most important,[25–28] but the bilayer motif was reported for VOPc on HOPG [29] and on Bi(111).[30]

In this work, we used a combined approach of STM and density functional theory (DFT) to study VOPc adsorption on Ag(100) with systematically increasing coverage from isolated molecules to few-layer thin films. At sub-monolayer coverage, O-up and O-down adsorption orientations can be distinguished by the contrast levels in STM images, with a population imbalance consistent with calculated adsorption energy differences. We find that the achiral VOPc molecules become chiral upon adsorption in the O-up orientation. As coverage is increased to ~1 ML, the molecules self-assemble into a commensurate (5x5)R+37° overlayer with respect to the Ag(100) surface. Unlike the mix of O-up and O-down seen at lower coverage, second layer VOPc molecules adsorb exclusively in the O-down orientation. Upon further deposition, subsequent layers are found to orient in an alternating O-up and O-down fashion, thereby establishing the bilayer motif. STM imaging of the bilayer surfaces reveals an additional long-range order that is moiré-like in appearance, but that cannot be attributed to calculated moiré lattices from the molecular and Ag(100) lattices. Chiral domains and grain boundaries are resolved in 2 ML and 4 ML regions, suggesting that the substrate-induced chirality may persist in film growth.

**Methods**

The Ag(100) surface was cleaned by repeated cycles of Ar$^+$ ion sputtering and annealing (~550 °C) in a ultra-high vacuum (UHV) chamber with a base pressure of ~1E$^{-10}$ mbar. VOPc molecules were degassed in an alumina coated crucible for 1hr at ~150 °C, as measured by a thermocouple attached to the side of the crucible. A low flux of molecules (~0.6 ML/min) was deposited onto the Ag substrate at room temperature by heating the crucible to 200 °C. Sublimation of VOPc was confirmed by a residual gas analyzer. Deposition was initially confirmed by Auger electron spectroscopy and then by direct STM imaging.

STM measurements were performed in two different systems: ~0.2 ML and ~3 ML coverage surfaces were studied at 78 K, while ~1 ML surfaces were studied at 5 K. 1 ML is defined as a single close-packed molecular layer that completely covers the substrate with the VOPc macrocycle planes oriented parallel to the surface. The results from the two systems are consistent, indicating that thermal adsorption processes (e.g. diffusion) below 78 K can be neglected. The STM measurements were performed with an electrochemically etched PtIr tip. Images have been colorized to emphasize relevant contrast using Gwyddion,[31] SpectraFox,[32] and WSxM.[33]

DFT calculations were performed using plane-wave basis and pseudopotentials as implemented in Quantum Espresso (V6.8 and V7.0).[34,35] We used projector augmented-wave (PAW) pseudopotentials from the PSLibrary.[36] The exchange correlation was approximated using the PBE functional,[37] and dispersive forces were treated using the revised VV10 method (rvv10).[38] We used rVV10 throughout this work as it is the most reliable dispersion correction for systems with mixed bonding environments.[38] All cells were built by creating suitable lateral supercells of the relaxed simple unit cells and padded in z-direction with ~1.5 nm of vacuum and decoupled from the periodic images in z-direction by using dipole correction. Cutoffs for the energy and charge density were chosen at 80 Ry and 640 Ry, respectively, and integration of the Brillouin zone was performed using only the Gamma point. In all calculations we applied a Hubbard U correction on the transition metal core of the molecules to improve the description of the localized 3d states and restores the correct orbital order in the case of VOPc.[39] We find a similar HOMO-LUMO gap of ~1.6 eV for any value of U>2.1 eV on the V 3d states, and we therefore used the value of U = 2.1 for the shown calculations. DFT-simulated STM images and were obtained using STMpw.[40]

**Results and Discussion**

Figure 1a shows an STM topography image of the Ag(100) surface with a low (~0.2 ML) coverage of VOPc molecules. The molecules image with nearly four-fold symmetry. The four lobes correspond to the four isoindole subunits which make up the Pc macrocycle (Figure 1a, inset). The vast majority of molecules fall into one of two distinct contrast levels which we attribute to two orientations: one in which the central O atom protrudes away from the surface (O-up, orange circle) and another in which the O atom points down towards the substrate (O-down, blue circle). For a small fraction of the molecules, we observe unstable imaging or rotation (those likely adsorbed on defect sites), or dissociated states such as VPc or Pc. Although the exact appearance of the VOPc molecules varies with tip termination and imaging conditions, the O-up and O-down orientations can be distinguished reliably by the relative apparent heights of their centers and lobes. The apparent height of the O-up molecules is lower (~129 pm) than O-down (~139 pm), and O-up molecules display articulated contrast at the center relative to the lobes, whereas the O-down molecules exhibit a more uniform height distribution. These assignments are consistent with prior STM studies of VOPc[25–27] and the isostructural MPc TiOPc.[19] From an analysis of 4363 molecules, O-up was found to account for 56±1% of the total, suggesting the O-up orientation is favored energetically. Further information on the atomic scale orientation of the molecules is revealed in a higher magnification STM image (Figure 1b). Although we cannot unambiguously determine the adsorption sites of the molecules from our data, we can identify the Ag(100) lattice directions from atomically resolved images (Figure 1b, inset). We find that the O-down molecules preferentially adsorb with lobes pointing parallel to the [011] direction, while the O-up molecules adsorb with lobes at ±27±1° to the [011] direction. Although this 27° angle is initially surprising given the square symmetry of both the molecule and the surface, this behavior is reminiscent of a prior study of CuPc on Ag(100) which suggested that the rotation is likely due to bonding between the ligand N atoms with underlying Ag atoms.[8]

These observations are consistent with our DFT calculations of isolated molecules. Our calculations suggest that O-down species tend to favor adsorption at hollow sites with a rotation of 0° relative to the [110] direction. Conversely, for the O-up configuration, molecules with rotations of both 15° and 30° are energetically allowed, with experimental images consistently showing isolated O-up molecules adsorbed at 30°. DFT-simulated STM images for both orientations are shown in Figure 1c,d. In agreement with experiment, the simulated images show more centralized contrast for O-up molecules, while the contrast variation is rather low across the center and lobes for the O-down molecules. Total energy calculations indicate a preference for O-up adsorption over O-down by 200 meV, in qualitative agreement with the observed population imbalance.

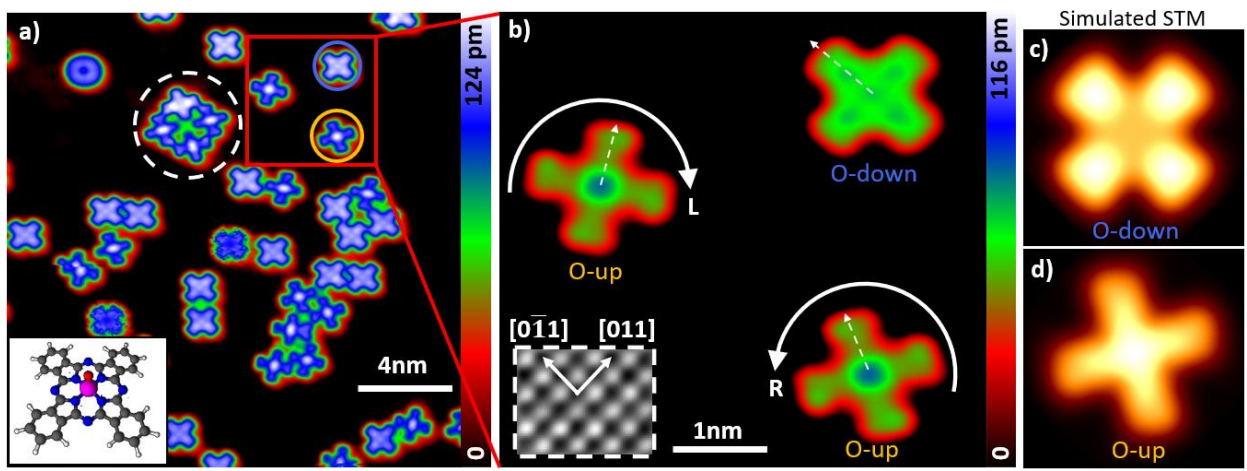

**Figure 1: Sub-monolayer VOPc on Ag(100).** (a) STM topographic images showing the coexistence of O-down (blue circle) and O-up (orange circle) molecules. White circle: cluster of four molecules forming a void with right-handed chirality. The O-down molecule has rotated to align with the other three molecules. Inset: Schematic of VOPc molecule. (b) Close up of three molecules in the red box in (a). The two O-up molecules exhibit opposite chirality, while the O-down molecule is achiral. Color scale has been chosen to emphasize differences in chirality. Inset: atomic resolution STM image showing Ag lattice directions (white arrows). (c,d) DFT-simulated STM images of O-down (c) and O-up (d) molecules with same Ag lattice directions as in (a/b). STM imaging conditions: (a,b) (-1.5 V, 300 pA); atomic resolution inset: (+10 mV, 1nA)

Close inspection of Figure 1b reveals a chiral structure for O-up molecules, while O-down molecules are achiral. For O-up, each lobe is slightly taller on one of its edges, causing the molecule to take on a windmill appearance in either the clockwise (L) or counter-clockwise (R) sense. This observation is consistent with the simulated image in Figure 1d. We observe statistically equal numbers of R- and L- enantiomers in our STM images, consistent with the achiral Ag(100) surface. This chirality is apparent only when imaging with negative sample bias (filled states) which may be attributed to distortion of the highest occupied molecular orbital as inferred from similar surface-induced chirality of CuPc on Ag(100).[8]

Chirality also emerges at the organizational level due to intermolecular interactions in self-assembled structures. For example, the four-molecule cluster shown in Figure 1a (white dashed circle) has a void with right-handed chirality. The three O-up constituents of this cluster are similarly right-handed, whereas the single O-down molecule in the cluster has deviated from its preferred rotation angle due to the influence of its neighbors. This observation suggests that the intramolecular chirality of individual adsorbed molecules may set a preference for the organizational chirality of clusters and monolayers. Indeed, the correlation between intra- and inter-molecular chirality has been previously reported in CuPc[8] and TiOPc.[19] In the former case, it was found that attractive VdW forces between molecules favor homochiral domains with the same organizational chirality as their constituent enantiomers.

To further study the competition between molecule-substrate and intermolecular interactions, Figure 2 shows STM images with varying coverage up to ~1 ML. At 0.95 ML average coverage, VOPc molecules begin to assemble into larger clusters (Figure 2a). The close-up in Figure 2b reveals that the cluster of molecules has formed a mixed unit cell of alternating O-up and O-down molecules with a nearest neighbor separation of 1.25 nm. In this cluster, the molecules generally maintain the same in-plane rotations as in the isolated molecule case. This indicates that interaction with the Ag surface still wins out over intermolecular forces. However, as coverage is slightly increased to 1.02 ML (Figure 2c) the molecules self-assemble into a continuous film in which every molecule shares a common in-plane rotation that matches the preferred angle of the isolated O-up molecules. Both orientations are distributed randomly within the film, with the O-down species being slightly more prevalent (60±2%). This imbalance is in contrast to the isolated molecule case where O-up was more common. This finding is in qualitative agreement with our DFT calculations which show a hierarchy of stability: mixed (most stable) > O-down > O-up (least stable) for ML films. The 1.02 ML nearest neighbor distance

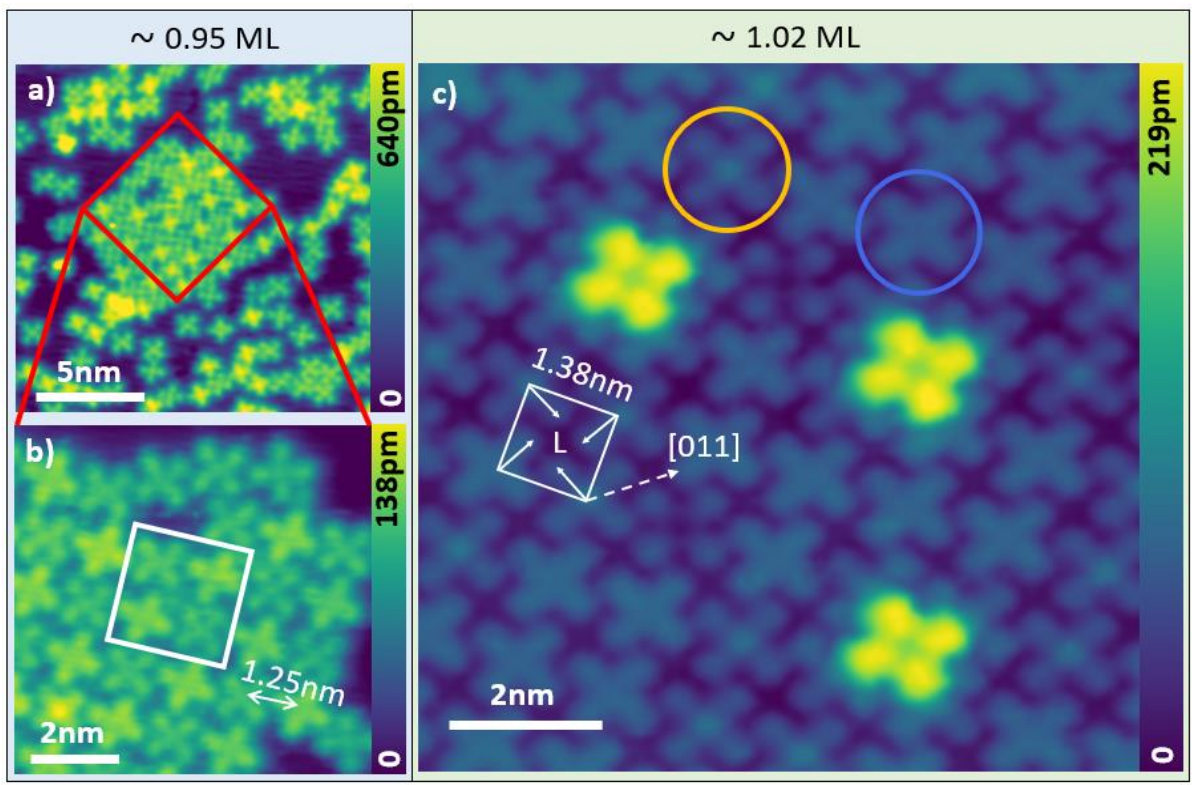

**Figure 2: 1 ML VOPc on Ag(100).** a) STM image of 0.95 ML VOPc showing clustering of O-up and O-down molecules. b) Close-up of the red-boxed area in (a) revealing a close-packed mixed unit cell of O-up and O-down molecules. c) STM image of 1.02 ML coverage showing self-assembly of VOPc into a square lattice with an increased inter-molecule spacing compared to the cluster in (b). O-up (orange circle) and O-down (blue circle) molecules are both present in the film, but 2$^{nd}$ layer molecules adsorbed atop the first layer are uniformly O-down. STM imaging conditions: (a) (+1 V, 50 pA); (b) (+0.1 V, 100 pA); (c) (+0.5 V, 100 pA)

increases compared to the 0.95 ML case to a value of 1.38 nm, which facilitates commensurate (5x5)R+37° ordering with respect to the Ag(100) surface. The transition from incommensurate clusters to a commensurate film suggests that molecule-substrate interactions outcompete intermolecular forces when the monolayer domain size becomes sufficiently large. Consistent with an increasing influence of intermolecular interactions at higher coverage, Figure 2c shows that second-layer VOPc molecules all appear identical and have dark centers, suggestive of an O-down orientation. The square overlay in Figure 2c highlights the left-handed organizational chirality of this region. This is demonstrated by drawing arrows from the VO centers along the axes of their lobes inside the square unit cell.

To further explore the orientation of second layer VOPc, we deposited a higher coverage (~3 ML) onto the sample (Figure 3a). We find that the surface is dominated by bilayer regions (2 ML, blue contrast) with sizeable 4 ML (orange contrast) islands, and only small regions of 3 ML and 5 ML VOPc. These assignments are supported by analysis of line profiles as in Figure 3b, which reveal a Ag(100) substrate step (204pm) between adjacent 2ML terraces, followed by a 311 pm step up to a small 3ML region, and a 304 pm step up to a 4ML region. As there were no 1 ML or exposed Ag regions, the parity assignment of the layers follows from consideration of layer-dependent contrast as we discuss below.

Focusing on the 2 ML surface, two length scales can be distinguished both from the real space image (Figure 3c), and the corresponding 2D-FFT (Figure 3d). First, a square lattice with spacing ~1.36 nm is observed, which matches the 1 ML surface (Figure 2c) to within experimental uncertainty. However, unlike the 1 ML surface where a mix of O-up/O-down molecules are observed, the uniform molecular contrast in Figure 3c indicates that molecules in the second layer are all oriented the same way. We assign this orientation to be O-down, based on a dark-center contrast similar to the admolecules in Figure 2c and the O-down molecules in Figure 1a. Furthermore, these data resemble STM images of bilayer VOPc on Ag(111)[25] and Bi(111)[30], which also terminate with O-down molecules. However, we observe an additional long-range ordering (LRO) in Figure 3c that was not previously reported on other surfaces. The LRO consists of a slightly warped square lattice with an average spacing of ~4.5nm, and a 49° rotation relative to the molecular lattice (blue square in Figure 3d). Though reminiscent of moiré lattices in 2D materials, we confirmed that the calculated spacing cannot arise from any rotations of the molecular layers with respect to each other or the Ag(100) surface, and may instead be due to structural warping of the bilayers as discussed further below.

In contrast to the even-layer surfaces, STM images of 3 ML regions (e.g., Figure 3e) and 5 ML regions show bright center contrast relative to the lobes, consistent with the O-up orientation. LRO is absent from these odd-layer surfaces. Two morphologies of 3 ML molecules were observed: free-standing tetramer clusters (dashed circle) and self-assembled islands. Each tetramer is centered on a molecule in the layer below, but is rotated by 37° with respect to the underlying molecular lattice. In contrast, the larger self-assembled islands are aligned with the underlying bilayer, and also show the same lattice constant. We found that such odd-layer islands were relatively rare and small compared to 2 ML and 4 ML regions, suggesting a strong preference for the bilayer motif. This motif is consistent with the bulk crystal structure,[41] and previous STM studies of VOPc[29,30] and other nonplanar MPcs.[15,42,43]

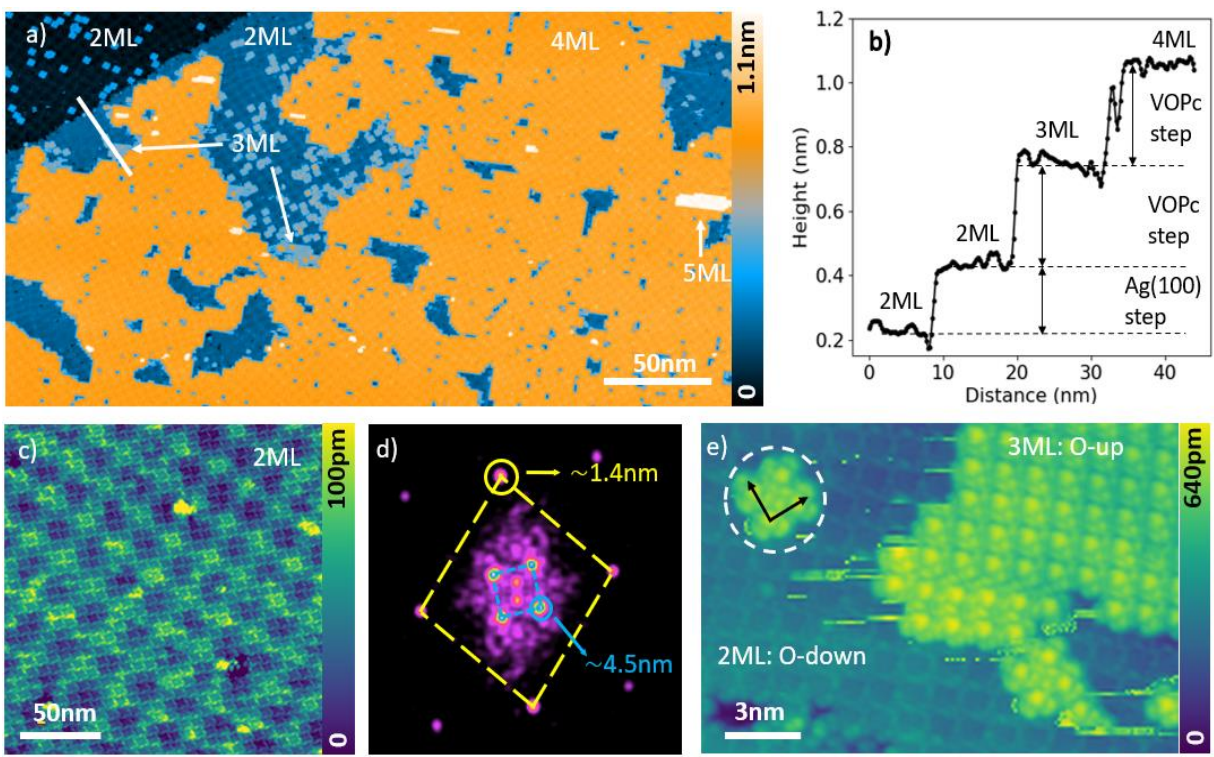

**Figure 3: Few-ML VOPc on Ag(100).** a) Region showing coexistence of 2 ML (blue) and 4 ML (orange) surface terminations. 3 ML molecules can be seen either as isolated tetramers, or small islands at the periphery of 4 ML islands. b) Profile corresponding to linecut in (a). Steps corresponding to the Ag(100) substrate and VOPc bilayer are labeled. c) Close-up of 2 ML region. A square lattice corresponding to the VOPc lattice can be discerned, along with long range ordering (LRO). d) 2d-FFT of image in (c). Yellow and blue squares correspond to the molecular lattice and LRO respectively. e) Close-up of 3ML island exhibiting O-up contrast. The 'scratchiness' near the edge of the island in Fig. 3e is likely due to tip-induced motion of mobile molecules near the island edge. STM imaging conditions: (a) (+1.8 V, 65 pA); (c) (+1.6 V, 100 pA); (e) (+2.5 V, 43 pA)

Figure 4a shows a high resolution STM image of a VOPc bilayer surface which allows a better view of the intramolecular contrast associated with the O-down orientation. The single molecule vacancy near the bottom of the image allows us to unambiguously register a molecular overlay. As illustrated the overlay, the molecular lattice is rotated by 37±1° with respect to the [011] direction. To within experimental uncertainty, this is consistent with a (5x5)R+37° overlayer, as we first noted for the 1ML coverage in Figure 2. As evidenced by the lobes of individual molecules, the VOPc molecules are rotated by 27±1° relative to the [011] direction, consistent with the angle observed in the 1ML case.

We find good agreement between these observations and our DFT calculations of the bilayer structure. The lowest energy structure is obtained by shifting and slightly warping the top layer of the structure so that O-up and O-down groups are offset and there is a net dipolar attraction associated with the oppositely oriented VO groups in the bottom and top layers (Figure 4b). Although we cannot determine the lateral offset between the first and second ML in our STM images, we measure a similar offset between the third and fourth ML in STM images taken near edges of 4ML islands. We note that the O-down termination in the DFT-simulated STM image (Figure 4a, inset) agrees well with the observed contrast. The predicted rotation of the molecules relative to the lattice vectors is also in agreement with experiment. Lastly, optimized bilayer structure involves a degree of out-of-plane warping, which can be seen in the lateral views in Figure 4b. The predicted warping of the bilayer provides a possible explanation of the LRO (which is observed only on even layers) as well as its distortion near step edges and grain boundaries.

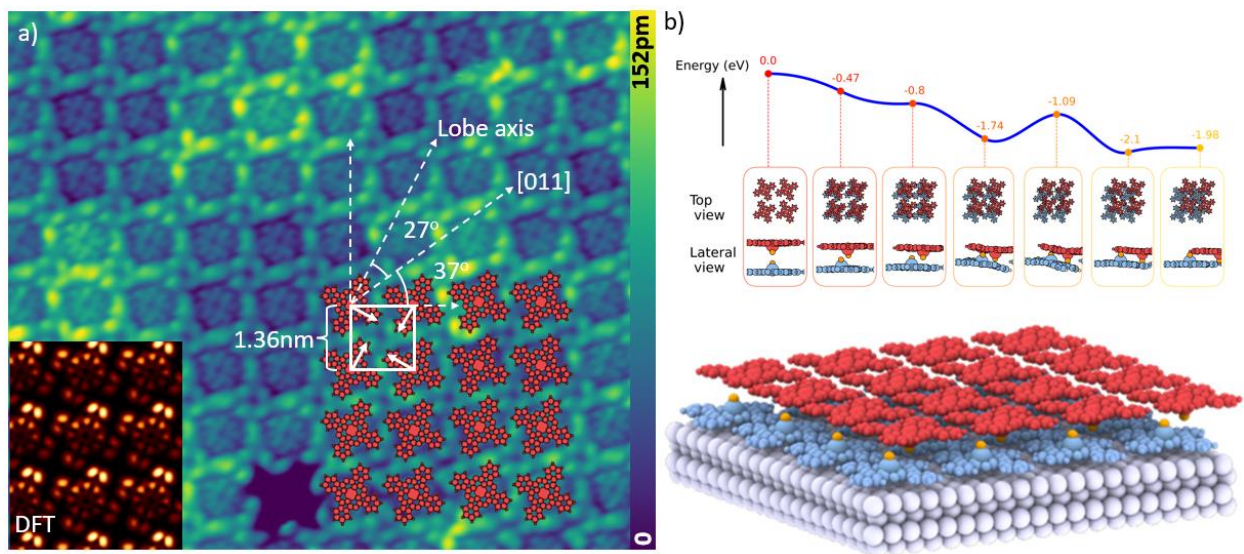

**Figure 4: Detailed comparison of STM and DFT of bilayer surface.** a) 2 ML surface with overlay revealing the packing structure of VOPc molecules. Dashed arrows denote the directions of the molecular lattice, substrate lattice, and lobe axes, while the solid arrows convey the organizational chirality. Inset: DFT simulated STM image of bilayer surface. b) (top) Plot of energy of bilayer system as a function of lateral offset between O-up and O-down layers. (bottom) 3D rendering of lowest energy bilayer configuration. STM imaging conditions: (a) (1.4 V, 100 pA). Image was corrected for thermal drift.

We conclude by returning to the notion of organizational chirality introduced in Figure 1. We find this chirality persists up through the multi-layer coverage. As the two organizational chiral domains are energetically equivalent, VOPc islands may nucleate with either chirality, and thus sometimes meet at a domain boundary as we show in Figure 5a. In this region, we identify distinct chiralities (labeled L and R) on either side of the boundary (dashed line), with a small achiral region (labeled A) next to the boundary. By adding arrows corresponding to the lattice vectors in each region, we see that they are different on either side of the boundary. While the L region corresponds to the previously discussed (5x5)R+37° overlayer, the R region corresponds to a (5x5)R-37° overlayer. While both overlayers are commensurate with the Ag(100) substrate, the mismatch between their lattice vectors prevents them from forming a single contiguous lattice. The lattice vector directions provide a quick method for spotting the chiral domains in larger-scale images, as demonstrated by the green and yellow axes overlaid in Figure 5b. The coexistence of the two domains in this image is confirmed by the FFT in Figure 5c which shows two sets of peaks (green and yellow squares). Figure 5b indicates that the chiral domains don't persist across substrate steps (red dotted line), but do continue in layer stacking. For example, the 4 ML island (orange) in Figure 5b exhibits like-chirality with the 2 ML surface underneath it. From an analysis of >30 unique 4 ML islands, we found no exceptions to this preference for like-chirality of subsequent bilayers. This suggests that the induced chirality associated with the Ag(100) interface would likely persist even in thicker films.

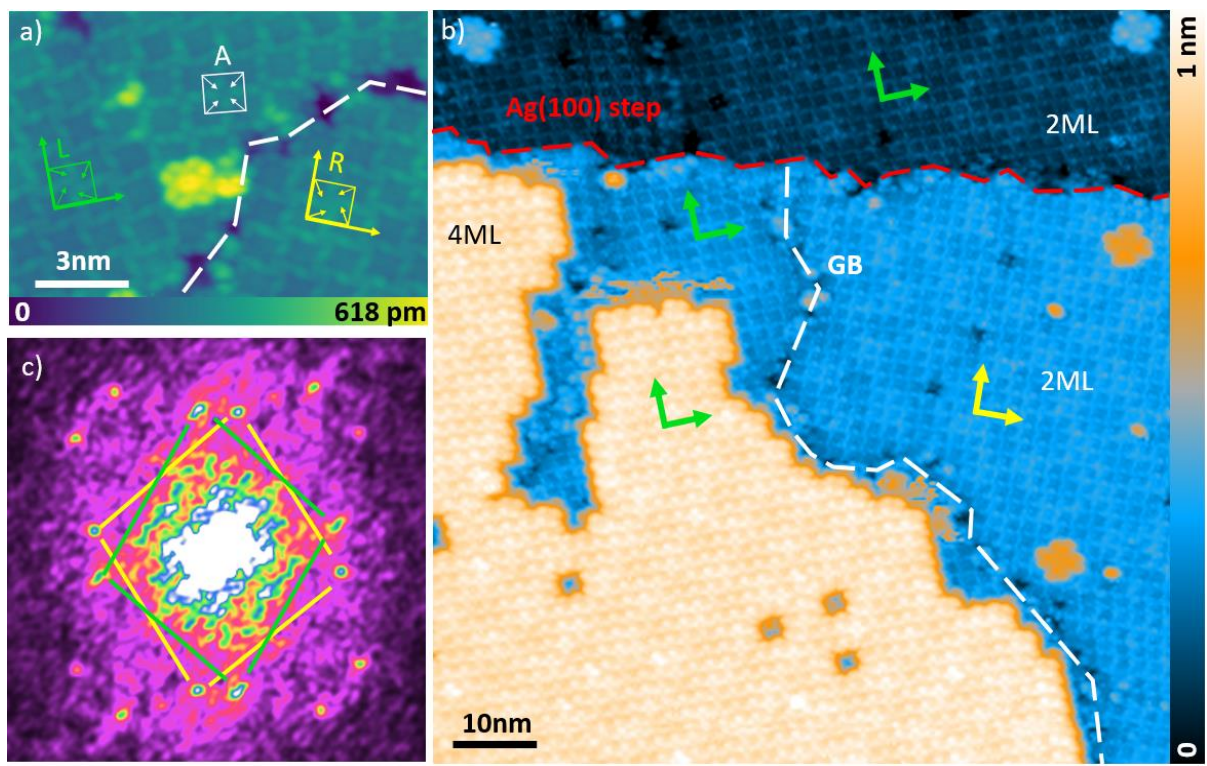

**Figure 5: Boundaries between chiral domains.** (a) High-resolution STM image of grain boundary (GB). Molecules on the left (right) side of the GB have left-handed (right-handed) organizational chirality, and lattice vectors oriented at +(-)37° with respect to the [110] direction. Near the boundary, there is an achiral organization. (b) STM image showing GB near a 4 ML island. The organizational chirality of the 2 ML (blue) and 4ML (orange) regions is denoted by the colors of the lattice vectors (green and yellow). (c) 2D-FFT of image in (b). The arrangement of four spots marked by the green (yellow) overlay corresponds to the left-handed (right-handed) chiral domain. Deviations from square symmetry may be due to drift and/or warping near GBs. STM imaging conditions: (a) (2 V, 20 pA); (b) (2.5 V, 100 pA)

**Acknowledgements**

This work was supported by funding from the National Science Foundation (NSF) QII-TAQS under award No. MPS-1936219, and by the Asian Office of Aerospace Research and Development (AOARD) (FA2386-20-1-4052).